\documentclass[12pt,preprint]{aastex} % one-column preprint format

\usepackage{graphicx}
\usepackage[twocolumn]{emulateapj5}

\newcommand{\sameauthor}{\underbar{\qquad\qquad}.}
\newcommand{\halpha}{H$\alpha$}
\newcommand{\hbeta}{H$\beta$}
\newcommand{\hgamma}{H$\gamma$}
\newcommand{\hdelta}{H$\delta$}
\newcommand{\vher}{V884~Her} % V884 Her
\newcommand{\kms}{km~s$^{-1}$}

\submitted{To appear in {\it The Astrophysical Journal}, Part 1}
\shortauthors{SCHMIDT, FERRARIO, WICKRAMASINGHE, SMITH}
\shorttitle{CYCLOTRON FUNDAMENTAL EXPOSED}

\begin{document}

\title{THE CYCLOTRON FUNDAMENTAL EXPOSED IN THE HIGH-FIELD MAGNETIC VARIABLE V884~HER}

\author{Gary D. Schmidt\altaffilmark{1,2}, Lilia Ferrario\altaffilmark{2},
D. T. Wickramasinghe\altaffilmark{2}, and Paul S. Smith\altaffilmark{1}}
\email{gschmidt@as.arizona.edu, dayal@maths.anu.edu.au, lilia@maths@anu.edu.au,
psmith@as.arizona.edu}

\altaffiltext{1}{Steward Observatory, The University of Arizona, Tucson, AZ 85721}
\altaffiltext{2}{Dept. of Mathematics, The Australian National University, Canberra,
ACT 0200 Australia}

\begin{abstract}

High-quality phase-resolved optical spectropolarimetry is presented for the
magnetic cataclysmic variable \vher.  The overall circular polarization during
active accretion states is low and only slightly variable in the range
$5000-8000$~\AA. However, the polarization is highly structured with
wavelength, showing very broad polarization humps, narrow features that are
associated with weak absorption lines in the total spectral flux, and sharp
reversals across each major emission line.  The polarization reversals arise
from Zeeman splitting in the funnel gas in a longitudinal magnetic field
$B\sim30$~kG.  The set of narrow, polarized absorption features matches the
Zeeman pattern of hydrogen for a nearly uniform magnetic field of $B=150$~MG,
indicating that the features are ``halo'' absorption lines formed in a
relatively cool reversing layer above the shock.  With this identification, the
broad polarization humps centered near 7150~\AA\ and below 4000~\AA\ are
assigned to cyclotron emission from the fundamental and first harmonic
($n=2$), respectively. \vher\ is only the second AM Her system known with a
field exceeding 100~MG, and the first case in which the cyclotron fundamental
has been directly observed from a magnetic white dwarf.

\end{abstract}


\section{Introduction}

The fate of cataclysmic variables containing the most highly magnetic primary
stars has long been a mystery.  Though single magnetic white dwarfs are
distributed continuously up to nearly 10$^9$~G (e.g., Wickramasinghe \&
Ferrario 2000), some 3 dozen AM Her-type variables (``polars'') were measured
to have polar field strengths $B_p<80$~MG (MG $=10^6$~G) before the first
high-field example, AR~UMa, was discovered at $B_p=230$~MG (Schmidt et al.
1996).  The observed characteristics of AR~UMa raised suspicions that the
high-$B$ examples might violate some of the hallmarks of the class: A very
large flux ratio of soft to hard X-rays suggests that the accreting gas
penetrates deeply into the photosphere, avoiding a standoff shock (e.g.,
Stockman 1988). As a result, the optical circular polarization of AR~UMa
during an active accretion state is nearly nil (Schmidt et al. 1996). The
magnetic white dwarf is also twisted in azimuth so that the magnetic axis is
nearly orthogonal to the stellar line of centers. Finally, AR~UMa has spent
the majority of the past several years in a state of inactivity, with
accretion episodes lasting at most a few months (Schmidt et al. 1999). The
implication is that a successful quest for high-$B$ cataclysmic variables
(CVs) may require broadening our definition of what constitutes a magnetic
accretion binary.

One clue in the search is the trend toward increasing soft X-ray ``excess''
with increasing magnetic field strength (Ramsay et al. 1994), apparently
caused by the progressive reduction in shock height and burying of
high-density portions of the accretion flow.  The most extreme example in this
regard is \vher\ (=RX~J1802.1+1804), an optically bright ($m_V\sim14-16$) but
very low-polarization system ($v\lesssim4\%$; Szkody et al. 1995) that has
been found to have a soft-to-hard X-ray flux ratio $F_{\rm SX}/F_{\rm
HX}\sim10^3-10^4$ (Greiner, Remillard, \& Motch 1998; Ishida et al. 1998;
Szkody et al. 1999). The orbital period of \vher\ is 113.01~min (1.8835~hr;
Greiner et al. 1998) and characterized by periodic dips in the X-ray light
curve whose depth shows strong energy dependence indicative of self-eclipse by
the gas stream. Similar features are present in the EUV (Hastings et al.
1999), but the optical light curve is less well-behaved. There is no detection
yet of either stellar component, a fact that prevents a reliable estimate of
distance and of the orbital location of the secondary at the time of the X-ray
dip. Some variation in accretion activity is implied by $1-2$~mag changes in
optical brightness with epoch, but thus far the binary has shown nothing
resembling an ``off'' accretion state, and Hastings et al. (1999) conclude
that high-state emission sources dominate even when $m_V\sim16$. The system
geometry is uncertain, but indications are of a rather low inclination
($i\sim45\arcdeg$) and small magnetic colatitude ($\delta\sim15\arcdeg$) with
a short ballistic section to the accretion stream (Greiner et al. 1998;
Hastings et al. 1999).

In this paper we present the results of phase-resolved optical
spectrophotometry, polarimetry, and spectropolarimetry of \vher\ that reveal a
circular polarization spectrum far more structured than any yet seen from a
magnetic CV.  Very broad polarization features are interpreted as emission in
the cyclotron fundamental and first harmonic in a magnetic field
$B\sim115-130$~MG, while a number of narrow features in polarized flux are
successfully modeled by hydrogen Zeeman features in a 150~MG accretion
``halo'' above the shock.  \vher\ reveals an important and heretofore
unobserved region of the spectrum of a magnetic accretion binary surrounding
the cyclotron fundamental.

\section{Observations and Results}

The bulk of the data reported here was acquired with the CCD
Spectropolarimeter (Schmidt, Stockman, \& Smith 1992) operating on the Steward
Observatory 2.3~m Bok telescope atop Kitt Peak.  The instrument was configured
for circular polarimetry over a full order of the grating,
$\lambda\lambda4000-8000$, providing a spectral resolution of $\sim$12~\AA.
All or most of an orbital period was covered on each of 3 epochs during the
period $1998-2000$, with waveplate sequences limited to $\sim$10~min in length
to achieve phase resolution of $\Delta\varphi\le0.1$. The system
brightness\footnote{These continuum magnitudes will register somewhat fainter
than broadband measurements because of the presence of emission lines in the
filter bands.} and degree of polarization from these measurements are
summarized in Table~1, together with the range in orbital phase, computed from
the X-ray dip ephemeris of Greiner et al. (1998):
\begin{equation}
{\rm HJD}=2449242.3124(21) + 0.07847977(11)\,E~.
\end{equation}
Additional single spectra were obtained as indicated in Table~1 to monitor the
state of activity, and a full orbit of ``white-light''
($\lambda\lambda3200-8600$) linear polarimetry with phase resolution
$\Delta\varphi=0.025$ was obtained with a filter polarimeter and the Steward
Observatory 1.5~m telescope on Mt. Lemmon.  The latter data reveal no
convincing detection of linear polarization above a 3$\sigma$ uncertainty
level of $P=2$\% for any sample during the series, such as might occur from
the beaming pattern of high-harmonic cyclotron emission.  Moreover, the
orbit-averaged result is only marginally significant at $P=0.30\pm0.11\%$;
$\theta=164\arcdeg\pm11\arcdeg$. A brightness modulation of amplitude 0.35~mag
is, however, apparent in the detected count rate at the 1.88~hr orbital
period.

The various observing runs sampled somewhat different levels of system
activity as indicated by the $\sim$1~mag spread in the visual magnitudes in
Table~1, but all of the spectropolarimetry show the same fundamental features:
a strong emission-line spectrum superposed on a blue continuum that steepens
in $F_\lambda$ below 5000~\AA\ and shows no sign of late-type stellar features
at red wavelengths. Emission-line ratios are typical for an AM Her system in an
active accretion state, with $F_{{\rm H}\alpha} \approx F_{{\rm H}\beta}
\approx F_{{\rm H}\gamma} \approx F_{\lambda4686}$. The circular polarization
is rather weak for a synchronized system, $|v|\lesssim5\%$ over the region
$5000-8000$~\AA. However, a very broad ($\Delta\lambda\sim1000$~\AA) negative
polarization bump is centered near $\lambda$7300 and negative polarization
sets in also for $\lambda\lesssim5700$~\AA.  The polarization increases
sharply to generally positive values around $\lambda4500$ and shows a great
deal of structure.  Finally, several ``narrow'' features,
$\Delta\lambda\sim100$~\AA, are scattered through the spectrum. These basic
characteristics are evident in the orbit-averaged results from 1999 Sep. 14
shown as Figure~1.

\subsection{Time Dependence}

Despite the differences in activity level, the spectrum-summed circular
polarization is negative at all 3 epochs, as it was when observed by Szkody et
al. (1995). Only a partial orbit was acquired on 2000 Mar. 10, yet the
polarization in this brighter state appears only slightly reduced in absolute
value.  When the individual spectra are phased according to the ephemeris in
eq. (1), significant Doppler motion of the emission lines is apparent,
reflecting the orbital modulation noted in previous spectroscopy. To study the
polarization at the highest signal-to-noise ratio (S/N), this variation was
removed from both the total flux and polarization spectra using a
half-amplitude of 150~\kms\ (e.g., Greiner et al. 1998, Hastings et al. 1999),
and the data from the multiple epochs were then coadded into phase bins. The
results are shown in Figure~2 as a phase sequence of circularly polarized
flux, $v \times F_\lambda$, and in Figure~3 as the phase dependence of circular
polarization in broad spectral bands selected to isolate major polarization
features. The grand sum is displayed together with the coadded spectral flux
at an expanded scale in Figure~4.

It is evident from Figures~2 and 3 that the polarized emission for
$\lambda\gtrsim5000$~\AA\ is largely constant throughout the orbit. Variations
are most apparent at the shortest wavelengths, where the polarized flux is
strongly positive soon after $\varphi=0$, switching to generally negative
during the opposite half of the orbit. Several of the narrow polarization
features noted above can be traced through the phase sequence in Figure~2,
and, as is evident in Figure~4, those which are uncontaminated by line
emission are accompanied by subtle depressions in the total flux.  The facts
that each of these absorption features exhibits a constant sign of
polarization and shows very little movement in wavelength indicate that, if
they are Zeeman in nature, only small changes in the mean magnetic field
strength can occur through the orbit. We will find that this conclusion has
important implications for the origin of the features.

\subsection{A Magnetized Accretion Funnel}

The circularly polarized flux displayed in Figure~2 and particularly in
Figure~4 shows sharp polarization reversals across each major emission line.
These are clear signatures of Zeeman splitting in the funnel gas.  The reality
of the features can be best appreciated from the individual panels of the
coadded sum shown as Figure~5, where the {\it sense\/} of the reversals ($-$
for $\lambda<\lambda_0$, $+$ for $\lambda>\lambda_0$) is seen to be common for
all, and the inflection in polarized flux occurs at line center for each
transition. The quality of the phase-binned polarimetry in Figure~2 is
sufficient to demonstrate that the sense of the splitting is also constant
throughout the orbit, however any estimate of phase-dependent variation in the
separation of the peaks would be severely limited by the spectral resolution
and S/N of the data.  We therefore measure the magnetic field present in the
emission-line region from the coadded panels of Figure~5.

The separation of the opposing polarization peaks ranges from 15~\AA\ for
\hdelta\ to 20~\AA\ for \halpha.  In weak fields such as this, the Zeeman
effect appears as a linear triplet for these transitions, and the $\sigma$
components separate from the central $\pi$ component at a rate of
$\pm20$~\AA~MG$^{-1}$ at \halpha, scaling as $\lambda_0^2$.  At face value
this would suggest a magnetic field strength of $\sim$500~kG (kG = 10$^3$~G).
However, this approach is invalid when the splitting is similar to the
spectral resolution of the data, a conclusion verified by the fact that the
field values so derived for the individual features increase monotonically
from $\sim$400~kG at \halpha\ to nearly 800~kG for the shortest-wavelength
lines. Moreover, this treatment ignores the fact that the emission line in
total flux possesses wings out to $\sim$1500~\kms\ ($\pm$25~\AA) from line
center, and that the entire profile probably experiences the magnetic field.
A proper approach must treat the line as two oppositely-polarized components
that are displaced slightly in wavelength, with the observed reversal due to
the tradeoff in dominance of one line wing over the other (e.g., Babcock
1947). The technique has been employed extensively in solar research, studies
of magnetic main-sequence stars, and was used to search for magnetic white
dwarfs to sensitivities of $\sim$1~kG (Schmidt \& Smith 1994; Schmidt \&
Grauer 1997). The instrument need only resolve the line profile, not the
Zeeman splitting itself.

A magnetic field value based on a circular polarization reversal measures the
mean longitudinal component of the field in the emission region, commonly
termed the ``effective'' field strength.  Using the analytical technique
described by Schmidt \& Smith (1994), we obtain measures of $B_{\rm eff}$ for
the individual strong lines as entered in Table~2.  Note that the dependence
of field strength on $\lambda$ has disappeared and because the polarization
reversals are due to the mismatch between line wings, the derived values are
much lower.  We conclude that the mean longitudinal magnetic field strength in
the funnel gas of \vher\ is $\sim$30~kG.  Arguments presented above regarding
the constancy of polarization sign in the lines suggest that orbital
variations do not carry the field lines through the plane of the sky, hence
that any phase dependence in field strength must be confined to the
approximate range $0\lesssim B_{\rm eff} \lesssim60$~kG. Assuming radial field
lines in the line-emitting portion of the funnel, this bounds
$i+\delta\le90\arcdeg$, the same limit as is placed by the presence of X-rays
through the entire orbital cycle (Greiner et al. 1998).

\section{Clues to a High Magnetic Field}

In attempting to understand the complex spectrum of broad and narrow
polarization features in \vher, we begin by looking to the basic properties of
the system for clues to the appropriate field strength regime.

First is the correlation between X-ray flux ratio $F_{\rm SX}/F_{\rm HX}$ and
magnetic field strength that was noted in \S1 as a motivation for this study.
The ratio of $>$1000 measured for \vher\ exceeds that on any other magnetic CV
(even AR~UMa only measures $\sim$60; Szkody et al. 1999), and for reasonable
extrapolations of Ramsay et al.'s (1994) correlation, suggests a polar field
$B_p\gtrsim80$~MG.

A second indicator for a high magnetic field is our detection of Zeeman
splitting in the funnel emission lines.  The only previous instance where this
has been observed is the $B_p=230$~MG system AR~UMa, where polarization
reversals with a separation of up to 30~\AA\ were found to form curious,
entwined patterns through the orbital cycle (Schmidt et al. 1999).  The basic
origin of that behavior is understood to be the transfer of polarized
radiation in optically-thick, magnetized streamlines (Ferrario,
Wickramasinghe, \& Schmidt 2000). The splitting in \vher\ retains a constant
sense throughout the orbit and amounts to $\sim{1\over2} - {2\over3}$ that in
AR~UMa. Assuming that the polarized line emission originates in the same
portion of the funnel in both systems and that all other factors are equal,
this suggests that the magnetic field on the white dwarf in \vher\ is in the
range $B_p=100-150$~MG.

A field of $100-150$~MG places the cyclotron fundamental at
$\lambda=1.1-0.7~\mu$m.  The optical portion of the spectrum would then sample
low-harmonic cyclotron emission, $n=1-3$.  Based on our understanding of the
lower-$B$ AM Her systems, we would not expect this emission to be strongly
beamed, so variability over the orbit will be small.  This is as observed in
\vher. Nevertheless, both the degree and variability in polarization will tend
to increase with increasing harmonic number (decreasing wavelength), also as
observed.

Thus, the properties of the impact region, the magnetic field strength in the
accreting gas, and the cyclotron emission characteristics all point toward a
high magnetic field on the white dwarf of \vher, $B_p\gtrsim100$~MG.  In the
following section we quantify this result through calculations of the hydrogen
Zeeman spectrum and polarization with the goal of modeling the pattern of
``narrow'' polarization features in our data.

\section{Zeeman-Split Halo Absorption Lines at B = 150 MG}

In Table~3 we list the locations of the circular polarization features that
are also visible in absorption and persist throughout the orbital cycle, as
well as a few weaker features that appear significant in the coadded data of
Figure~4. The fact that the features appear during what are clearly active
accretion states argues that they should be attributed to Zeeman-split
absorption in a relatively cool ``halo'' that overlies the bright continuum
source, and not photospheric features of the white dwarf. Halo absorption was
originally noted as a polarized \halpha\ component that was visible only
during the bright phase of ST~LMi (Schmidt, Stockman, \& Grandi 1983), but
rich patterns have now been detected in a number of systems, allowing
independent measures of the magnetic field strength near the shock (e.g.,
Wickramasinghe, Tuohy, \& Visvanathan 1987). The narrowness of the features in
\vher, coupled with the lack of significant wavelength shift with orbital
phase, is also indicative of an origin as halo Zeeman lines against a
localized bright spot, as opposed to the rotation of, e.g., an oblique dipolar
field pattern distributed over the surface of the star.

With the considerations of \S3 as a guide, we have found that the pattern of
features in \vher\ can be successfully interpreted as absorption by hydrogen
in an essentially uniform magnetic field of $B=150$~MG.  The quality of the
agreement is depicted in the top panel of Figure~4, where we include the
behavior of prominent components of the Balmer series for
$B=0\rightarrow150$~MG, taken from calculations of Forster et al. (1984),
Wunner et al. (1985), and Wunner, Geyer, \& Ruder (1987).  Identifications for
individual spectral features are given in Table~3.  Where strong signatures
appear in both polarization and spectral flux -- i.e. shortward of \halpha\ --
we consider the assignments certain.  The red features are more problematic,
being much weaker in polarization and confused by other spectral features. For
example, unpolarized line emission, like \ion{He}{1} $\lambda6678$ superposed
on the 2p1 -- 3s0 absorption, dilutes the degree of polarization but does not
directly affect polarized flux such as is displayed in Figure~4. Terrestrial
absorption bands, on the other hand, reduce the observed flux level and
therefore the measurement precision, but do not modify $v$(\%).  It should be
noted that the behavior shown in Figure~4 confirms our choice of a
uniform-field absorbing layer, since several of the transitions (esp. the
\halpha\ $\pi$ components at $\lambda\lambda$5516, 5778, 5930) are so
sensitive to magnetic field strength that they would be smeared to
invisibility if the spread were larger than $\Delta B \sim 10$\%.  Note also
that a tentative field strength estimate for \vher\ of $B_p\approx120$~MG
(Schmidt 1999) is superseded by this new measurement.

It is clear from Figure~4 that the prominent polarization features are all
positive-going excursions.  This includes purely $\sigma^+$ transitions like
the crossing of \hbeta\ components at $\lambda$4940, as well as the \halpha\
$\pi$ transitions in the range $5500-6000$~\AA.  In general, $\sigma^-$
components move so rapidly with $B$ that they are not observed in high-field
situations. What is surprising is that the observed $\sigma^+$ and $\pi$
components do not show detectable Doppler shifts with phase
($\Delta\lambda<\pm20$~\AA) as would be expected if these lines were formed in
free-falling pre-shock material.  In the lower-field systems EF Eri and V834
Cen, Doppler variations of $\Delta\lambda>\pm50$~\AA\ ($\Delta
v>\pm2000$~\kms) are observed as gas in the low-density parts of the accretion
flow is heated by hard X-rays (bremsstrahlung) emanating from shocks that are
formed at the base of high-density parts of the stream (Achilleos,
Wickramasinghe \& Wu 1992; Schwope \& Beuermann 1990).

The halo lines in \vher\ appear instead to be occurring in a virtually static
reversing layer. This difference may be related to the nature of the accretion
shock. The various possibilities have been summarized by Wickramasinghe \&
Ferrario (2000). At $B=150$~MG, the accretion shock is expected to be
cyclotron-cooling dominated. The shock height $h$ is expected to be very much
less than its lateral dimension $d$ so that the shock will be cooled from the
upper surface.  As a consequence, a steep temperature gradient is expected to
develop in the vertical direction, just as in a stellar atmosphere. We argue
below that the reversing layer that gives rise to the Balmer lines is most
probably composed of nearly stationary post-shock material that is heated by
the cyclotron radiation emanating from the accretion shock itself.  Static
reversing layers have been predicted for so-called ``bombardment'' solutions
(Woelk \& Beuermann 1992).  However, these are unlikely to apply in the
regions of high specific accretion rate that are required to provide the
underlying continuum against which the Balmer lines are formed.

In order to understand the physical processes that might be occurring in the
reversing layer, we first consider what we would expect from a photospheric
patch with a nearly uniform field strength on the surface of a magnetic white
dwarf. We set the effective temperature to $T_{\rm eff}=14,000$~K to ensure
that the reversing layer is cool enough to produce Balmer line absorption. The
mean field threading the patch is 150~MG, it has a spread in strength of
$\pm5$~MG, and is viewed at an angle of 50\arcdeg\ to the mean field
direction. The calculations are based on the magnetic white dwarf atmosphere
code developed by Wickramasinghe \& Martin (1979), and include
magnetobremsstrahlung opacity as in Pacholcyzk (1976). Magneto-optical effects
are included both in the lines and in the continuum, and for the individual
broadening of the Zeeman components, as described in Wickramasinghe (1995).

Results of the calculations are shown in Figure~6.  The model produces
circular polarization in the $\pi$ components, as observed in the data,
arising from the conversion of linear polarization to circular polarization
through Faraday mixing as the radiation propagates through the reversing
layer. The Faraday (or magneto-optical) effects involve the continuum since
they are seen in isolated $\pi$ components. These results strongly support the
identifications proposed in the previous section and listed in Table~3. Most
of the theoretically calculated transitions appear to be represented in the
data. Ignoring for the time being the broad, positive-going polarization bump
centered around $\lambda7150$, we see that the model reproduces the basic
characteristics of the observed polarization spectrum: a generally negative
level of circular polarization in the continuum and strong positively-directed
polarization features for each of the major halo Zeeman lines.  In good
correspondence to the data, the $\pi$ features tend toward, but do not exceed
$v=0$, while net positive circular polarization is computed for the strongest
$\sigma^+$ transitions. The model shows that the positively-polarized
$\sigma^+$ components that occur in the red part of the spectrum are much
weaker than the components at shorter wavelengths, and indeed they are also
very weak or absent altogether in the observations.

The photospheric model qualitatively explains the polarization properties of
the absorption lines but fails to provide an adequate description of the
observed continuum polarization. Both the narrow absorption feature predicted
at 7150~\AA\ and the associated broad polarization hump centered at this
wavelength result from photospheric absorption at the cyclotron fundamental.
Since a free electron gyrating in a magnetic field radiates the same sense of
circular polarization as a $\sigma^+$ line component, and we are dealing with
an absorption process, the predicted cyclotron feature shares the same
polarization characteristics as the $\sigma^+$ components. The width of the
cyclotron feature results from the field strength spread that has been allowed
for in the atmosphere calculation. The cyclotron absorption feature is,
however, not present in the data, and the data show a strong negative
polarization hump in the wavelength region $6700-8000$~\AA, the opposite sense
to what is calculated in the model. Furthermore, there are spectral regions in
the data where the polarization is essentially zero, which is again
inconsistent with the photospheric model. The shortcomings of the model in
regard to the continuum polarization suggest that although the reversing halo
is optically thick to the Balmer lines, it is optically thin to cyclotron
absorption and also to other continuum absorption processes such as bound-free
absorption. Indeed, the spectral intervals with very low polarization are
appropriate for continuum emission that originates from an optically-thick
underlying shock and passes through the reversing layer virtually unaffected.
The situation is therefore different from the photospheric model that we have
presented since in the model the radiative flux at large optical depths
becomes polarized as it flows through the atmosphere.

\section{A Cyclotron Origin for the Polarization Humps}

At $B_p=150$~MG, the cyclotron fundamental lies at 7150~\AA.  The existence of
the broad polarization hump around this wavelength cannot be coincidental, and
we identify it with emission at the fundamental. Because it is in emission,
the polarization has the opposite sign to that of the positively-polarized
$\sigma^+$ absorption features.  In addition, it appears asymmetric in
polarized flux (Figure~4), with a tail extending to long wavelengths.  Red
wings to cyclotron emission features result from the high-energy tail of the
relativistic Maxwellian electron velocity distribution (e.g., Wickramasinghe \&
Ferrario 2000).  The first cyclotron harmonic ($n=2$) would then be located
off the blue end of our spectrum near 3600~\AA, and we identify both the gently
increasing (negative) polarized flux as well as the rise in total spectral
flux as one moves shortward of 6000~\AA\ as the red tail of this peak.  In
isolation, the tail would presumably increase smoothly in strength and connect
with the strong negatively-polarized segment in the region $4050-4200$~\AA.

The polarization reversal between 4200~\AA\ and 5000~\AA\ could in principle
be a cyclotron feature from a second accretion pole of opposite polarity, but
the absence of another positively-polarized harmonic in the observed spectral
region would require that this be the fundamental, implying a field strength
at the second pole of $\sim$240~MG.  Moreover, the fact that both this hump
and the 7150~\AA\ feature are present throughout the entire orbit would
require both poles to be on our hemisphere continuously, an unlikely
configuration for standard dipolar field geometries. Instead, we point out the
confluence of Zeeman halo components from \hgamma\ and \hbeta\ that fall in
the region $\lambda=4000-5000$~\AA, only the strongest of which are indicated
in Figure~4. As we have seen, all of the $\pi$ and $\sigma^+$ transitions
yield positive features in circular polarization, and taken together could
produce the structured hump in this region.  The Zeeman features are seen
against the beamed higher-harmonic cyclotron emission, which is more sensitive
to viewing angle than the 7150~\AA\ fundamental, and we attribute the phase
dependence of polarized flux in the $4000-5000$~\AA\ region (Figure~2) to
variable cancellation as the white dwarf rotates. As a consistency check on
this explanation, we note from Figures~2 and 3 that when the cyclotron features
(filled circles and filled squares in Figure~3) are strongest,
$0.4\lesssim\varphi\lesssim0.8$, the $4000-4250$~\AA\ peak (open circles) is
strongly negative and the positive hump (open triangles) is relatively weak.
The converse is true when the cyclotron features are comparatively weak
($0.0\lesssim\varphi\lesssim0.2$).

Within the context of the modern model of profiled accretion shocks in AM Her
variables (e.g., Schmidt, Stockman \& Grandi 1986; Wickramasinghe \& Ferrario
1988; Wu \& Chanmugam 1988), the low-harmonic cyclotron emission from \vher\
presumably arises in a low-$\dot m$ tail of very low optical depth
($\Lambda\lesssim100$ to provide the observed polarized emission within the
harmonic) and comparatively large area. In attempting to model this region, we
note that the polarization of the Zeeman lines constrains the angle between
the line of sight through the halo and the field direction to be in the range
$0\arcdeg\lesssim\theta\lesssim60\arcdeg$ over the orbit.  The lack of
movement with phase of the 7150~\AA\ cyclotron feature places additional
restrictions on the range in $B$ and $\theta$ for the region producing the
low-harmonic emission. Of course, these constraints must be consistent with
limitations on $i$ and $\delta$ derived from our emission line
spectropolarimetry (\S2.2) and the light curve at higher energies (e.g.,
Greiner et al. 1998; Hastings et al. 1999).

The simplest of models used to calculate cyclotron spectra are
constant-$\Lambda$ models (Wickramasinghe and Meggitt 1985) that assume
uniform conditions within the shock but allow for optical depth effects. The
emergent intensity and polarization along a given pathlength are calculated as
a function of viewing angle $\theta$, electron temperature $T_e$, harmonic
number $n$, and optical depth parameter $\Lambda$.  We show in Figure~4 a
series of polarization spectra from models that have been constructed with
$T_e=5$~keV and $\Lambda=20$. The choice of $T_e$ is arbitrary, but the
unusually low value for $\Lambda$ ensures that the cyclotron fundamental is
marginally optically thick and polarized.  The sequence corresponds to viewing
angles $\theta=20\arcdeg,30\arcdeg,40\arcdeg,50\arcdeg$ relative to the field
direction and field strengths $B=115,120,125,130$~MG respectively.  In each
case, the field has been chosen so that the broad polarization feature
centered at 7150~\AA\ corresponds to the cyclotron fundamental.  The
calculations show that it should be possible to obtain qualitative agreement
with the observations using a structured shock model that encompasses a
spread in field strength and direction. For example, the 50\arcdeg\ model
exhibits a large excursion to negative polarization around 4000~\AA, very
similar to the observations during the $\varphi=0.4-1.0$ interval (Figure~2).
Conversely, the model for $\theta=20\arcdeg$ shows only a weak $n=2$ harmonic
feature, and the polarized flux would be dominated for $\lambda\lesssim5000$
by the positively polarized Zeeman lines, resembling the data for the
$\varphi=0.0-0.2$ period.  A small viewing angle with respect to the field at
this time is in agreement with the presence of an X-ray eclipse due to funnel
material that defines $\varphi=0$.

These models are, however, by no means unique. For the calculations presented,
the change in polarized flux between the first harmonic and the fundamental is
due purely to a change in viewing angle.  The polarized flux ratio is also
sensitive to $\Lambda$, and in a true structured shock the two variables are
unlikely to be independent.  A different choice for $T_e$ would also result in
a similar series of model fits, but with a different value of $\Lambda$.
Indeed, the frequency of a given cyclotron harmonic -- and therefore the
derived field strength -- depends on $B$, $T_e$, and $\theta$. For the 5~keV
models, the disparity between the Zeeman field of 150~MG, and the cyclotron
field of $115-130$~MG, is significant. Lower electron temperatures would bring
the two values into better agreement, but it is then difficult to obtain
significant polarized flux at the first harmonic ($n=2$). A higher viewing
angle would also decrease the discrepancy, but this would imply that the field
direction in the cyclotron region is significantly different from that of the
halo. For a centered dipole, the disparity between the halo Zeeman and
cyclotron field strengths implies an unlikely large geometrical extent to the
shock region ($\Delta\delta \sim 40\arcdeg$).  Again, this can be reduced if
one adopts a more complex field structure.  A dipole offset along its axis
toward the observer increases the field gradient along the stellar surface,
decreasing the corresponding length of the accretion arc.

Of course, the actual observed degree of polarization at any wavelength
depends on the importance of competing sources of light.  Between emission
harmonics, the accretion tail is optically thin and the light output is
presumably dominated by the optically thick, and thus unpolarized,
Rayleigh-Jeans emission from interior shock regions that have the
traditionally large values of $\Lambda$ ($\sim$$10^4-10^6$). This region of
the flow is what produces the strongly polarized and often well-defined
cyclotron emission peaks in the higher harmonics ($m\sim5-10$) that are
observed in the optical from primary stars with lower field strengths.  In
\vher, the corresponding spectral region lies in the UV. Because the net
polarization within the low-$\dot m$ cyclotron emission features only reaches
$v\sim5-10$\%, and the harmonic emission itself is likely strongly polarized,
the low-$\dot m$, low-$\Lambda$ tail may provide as little as 10\% of the
light at those wavelengths.

In closing this section, we note that a variety of alternative interpretations
of the polarization spectrum were considered, using traditional high-$\Lambda$
shocks.  These semi-quantitative models attempted to account for the
low-polarization spectral regions by assigning them to cyclotron emission
harmonics, while the strongly polarized intervals were imagined to lie between
harmonics.  The field strength in the cyclotron-emitting region would then be
$\sim$175~MG.  However, it was not possible to reproduce the polarization
reversal and variability below 5000~\AA\ without appealing to gross
temperature and/or field structure in the shock or to the appearance of a
second accretion pole of opposite polarity that is always in the visible
hemisphere (as in a quadrupolar field structure). For the sake of simplicity
and self-consistency, we prefer an explanation in terms of low-$\Lambda$
emission harmonics. With this insight, we are now able to successfully
interpret the polarization spectrum of the highest-field magnetic CV AR~UMa
(Ferrario et al. 2000).

\section{Conclusions}

As the only second AM Her variable with $B_p>100$~MG, \vher\ displays yet
another dimension to the phenomenon of magnetic accretion.  In contrast to
AR~UMa, whose accretion duty-cycle has been $<$25\% since it was first
observed optically in 1991 (Schmidt et al. 1999), \vher\ has never been
detected in a state of inactivity.  Our spectropolarimetry in a comparatively
low state, some 2~mag below historical maximum light, reveals a spectrum still
dominated by polarized cyclotron emission and a brilliant array of emission
lines from the accretion funnel.  The contrast in properties between the two
systems casts further doubt on the magnetic field having any significant
effect on mass loss from the companion star (see also Schmidt et al. 1999);
instead we conclude that the accretion rate is more likely driven by processes
which vary considerably from one object to the next and/or that are highly
variable on timescales of at least a decade.

Though not recognizable in total flux spectra, Zeeman splitting in the
emission lines has now been detected through circular spectropolarimetry in
both systems with $B>100$~MG, providing a potential means of identifying
high-field AM Her variables where direct identification of photospheric or halo
absorption features is not possible.  Field strengths measured from these
magnetic splittings confirm model results (Ferrario \& Wehrse 1999) indicating
that the bulk of the observed line emission in AM Her systems arises
$\sim$10~$R_{\rm wd}$ from the surface, where the product of the funnel
surface brightness and projected area is a maximum.

Less quantitative but possibly more useful as a hallmark of a high magnetic
field is the relation between soft X-ray excess and magnetic field strength.
Though variables such as accretion rate and blob spectrum undoubtedly play
roles, the values of $F_{\rm SX}/F_{\rm HX}$ measured for AR~UMa and \vher\
count among the highest ratios measured for AM Her binaries.  These two
systems provide strong confirmation of the expected reduction in standoff
height with increasing field strength, leading to eventual burying of the
shock. For AR~UMa, the resulting quenching of cyclotron emission appears to be
complete in the optical; for the somewhat lower-field system \vher, the
circular polarization is weak around the fundamental but increases with
decreasing wavelength.

In \vher, we have for the first time observed the cyclotron fundamental from a
magnetic white dwarf.  This came through the measurement of circular
polarization, not total flux, and the polarized emission appears to originate
in a low-$\dot m$ tail of the impact region.  This tail presumably represents
the first or last (or both!) gas to be stripped from the ballistic stream, a
characteristic that may contribute to the apparent discrepancy in field
strengths derived from the low emission harmonics {\it vs.\/} the halo Zeeman
lines. Optical depth effects in this low-$\Lambda$ portion of the shock may
also play a strong role.  We expect emission at the fundamental to be common
among AM Her systems, where it is accessible. The $B_p=92$~MG RX~J1007.5$-$2016
(Reinsch et al. 1999) is an excellent prospect, in view of the fact that the
$n=2$ and 3 harmonics are prominent in spectral flux.  Proper interpretation
of these features poses new challenges for understanding the physical
properties of cyclotron-cooled accretion shocks, an area not yet explored
fully in astrophysics. Calculations of the type presented by Woelk \&
Beuermann (1996) need to be extended to the high-field regime, allowing in
addition for 3-D shock structure.

\acknowledgments

G.D.S. is grateful for the hospitality and support of the Australian National
University and Mount Stromlo Observatory, where this project was completed
during a sabbatical leave. Studies of magnetic stars and stellar systems at
Steward Observatory are supported by NSF grant AST 97-30792.

\clearpage

\begin{deluxetable}{lcccc}
\tablenum{1}
\tablewidth{0pt}
\tablecaption{Log of Observations}
\tablehead{\colhead{UT Date} & \colhead{$\varphi$} & \colhead{Type} & \colhead{$m_V$ (mag)}
& \colhead{Polarization} \\
\colhead{(yyyymmdd.dddd)} & & & & }
\startdata
19980522.4616$-$.4689 & 0.60$-$0.69 & Cir. Spectropol. & 16.1        & $-1.83\%$ \\
19980923.1077$-$.2060 & 0.07$-$1.32 & Cir. Spectropol. & 16.2$-$16.6\tablenotemark{a} & $-1.32$ to +0.57\% \\
19990518.3515$-$.4376 & 0.11$-$1.25 & Cir. Spectropol. & 16.0$-$16.3\tablenotemark{a} & $-1.20$ to +0.14\% \\
19990914.1599$-$.2476 & 0.94$-$1.06 & Lin. Filter Pol. & $\cdots$    & $0.30\pm0.11\%~@~164\arcdeg$ \\
19991014.1495$-$.1568 & 0.05$-$0.14 & Spectroscopy     & 15.3        & $\cdots$ \\
20000310.4264$-$.4796 & 0.42$-$1.10 & Cir. Spectropol. & 15.7$-$15.9 & $-1.00$ to $-0.75$\% \\
\enddata
\tablenotetext{a}{Intermittent light clouds}
\end{deluxetable}

\begin{deluxetable}{lc}
\tablenum{2}
\tablewidth{0pt}
\tablecaption{Zeeman Splitting in the Emission Lines}
\tablehead{\colhead{Feature} & \colhead{$B_{\rm eff}$ (kG)} }
\startdata
\halpha & 25 \\
\hbeta & 32 \\
\ion{He}{2} $\lambda$4686 & 35 \\
\hgamma & 20 \\
\hdelta & 21 \\
\enddata
\end{deluxetable}

\vskip-3.truein
%\clearpage

\begin{deluxetable}{cc}
\tablenum{3}
\tablewidth{0pt}
\tablecaption{Polarized Halo Lines of Hydrogen}
\tablehead{\colhead{$\lambda$ (\AA)} & \colhead{Identification ($B$=150~MG)} }
\startdata
4296 & 2p$-$1 -- 3d0 ($\sigma^-$), 2s0 -- 4f0 ($\pi$), 2p1 -- 5s0 ($\sigma^+$) \\ % s-, pi, s+
4480 & 2p0 -- 3s0 ($\pi$), 2p0 -- 4d$-$1 ($\sigma^+$), 2p1 -- 5d0 ($\sigma^+$) \\ % pi, s+, s+
4940 & 2s0 -- 4f$-$1 ($\sigma^+$), 2p1 -- 4s0 ($\sigma^+$) \\ % s+, s+
5516 & 2p$-$1 -- 3d$-$1 ($\pi$) \\ % pi
5778 & 2p0 -- 3d0 ($\pi$) \\ % pi
5930 & 2s0 -- 3p0 ($\pi$) \\ % pi
~~~~~6645,6690 (?) & 2p1 -- 3s0 ($\sigma^+$) \\ % s+
~~~~~6847 (?) & 2s0 -- 3p$-$1 ($\sigma^+$) \\ % s+
\enddata
\end{deluxetable}

\clearpage

\begin{figure}%Figure 1
\vskip 2.5truein \includegraphics{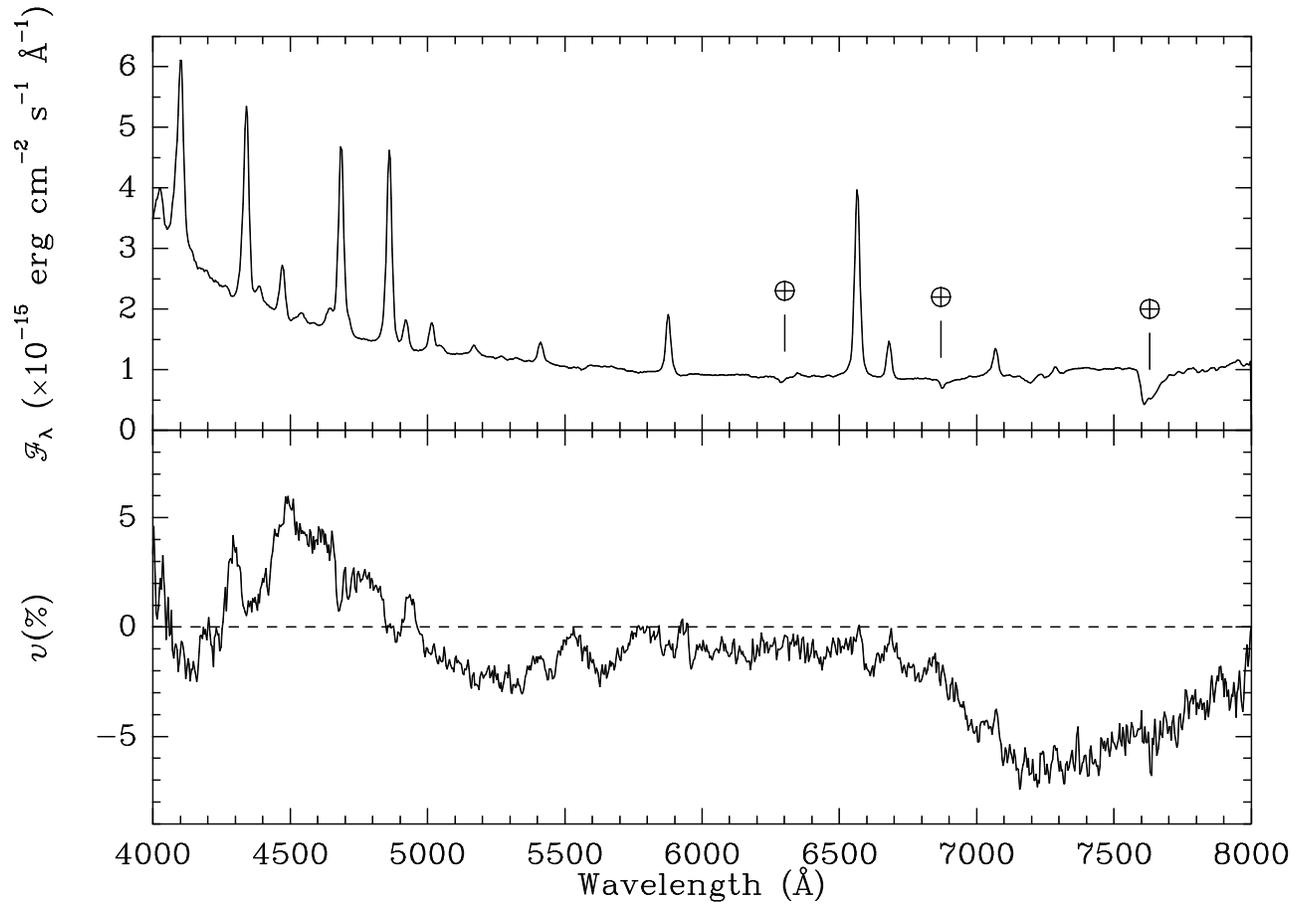}

\caption{Orbit-averaged total flux and circular polarization spectra of \vher\
from 1998 Sep. 14. Note the comparatively weak level of circular polarization
for a magnetic CV and the highly structured appearance, particularly for
$\lambda<5000$~\AA.  Terrestrial absorption features affecting the spectral
flux are noted.}
\end{figure}

\clearpage

\begin{figure}%Figure 2
\vskip 7.4truein
\includegraphics{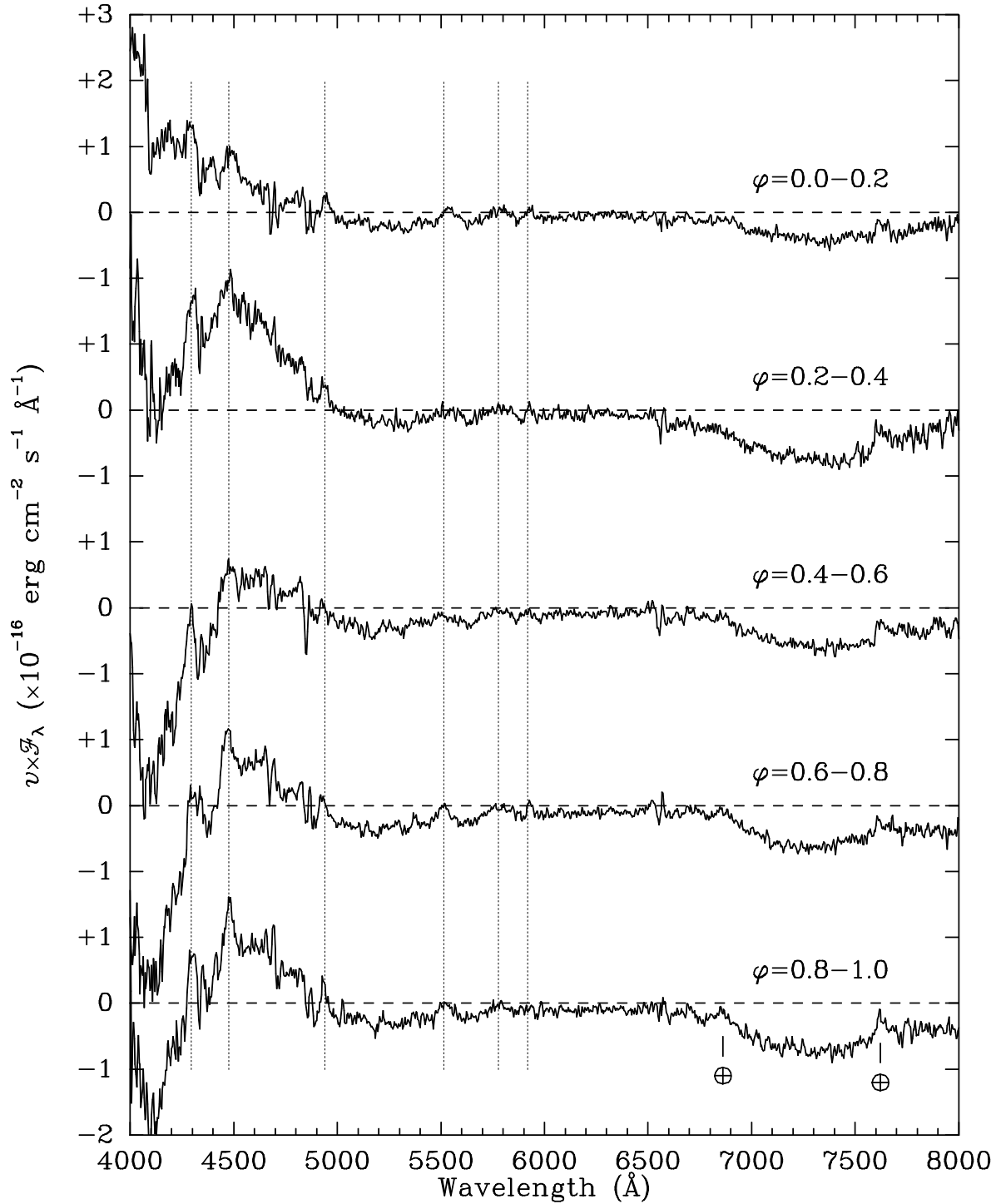}

\caption{Phase-binned circularly polarized flux of \vher\ averaged over all 3
epochs and with the orbital motion removed.  Variability is confined largely
to the bluest wavelengths.  Sharp polarization reversals in each of the major
emission lines \halpha, \hbeta, \ion{He}{2} $\lambda$4686, etc. signify Zeeman
splitting in the funnel.  Several narrow polarization features that remain
nearly stationary with phase are indicated by dotted lines.}
\end{figure}

\clearpage

\begin{figure}%Figure 3
\vskip 2.5truein
\includegraphics{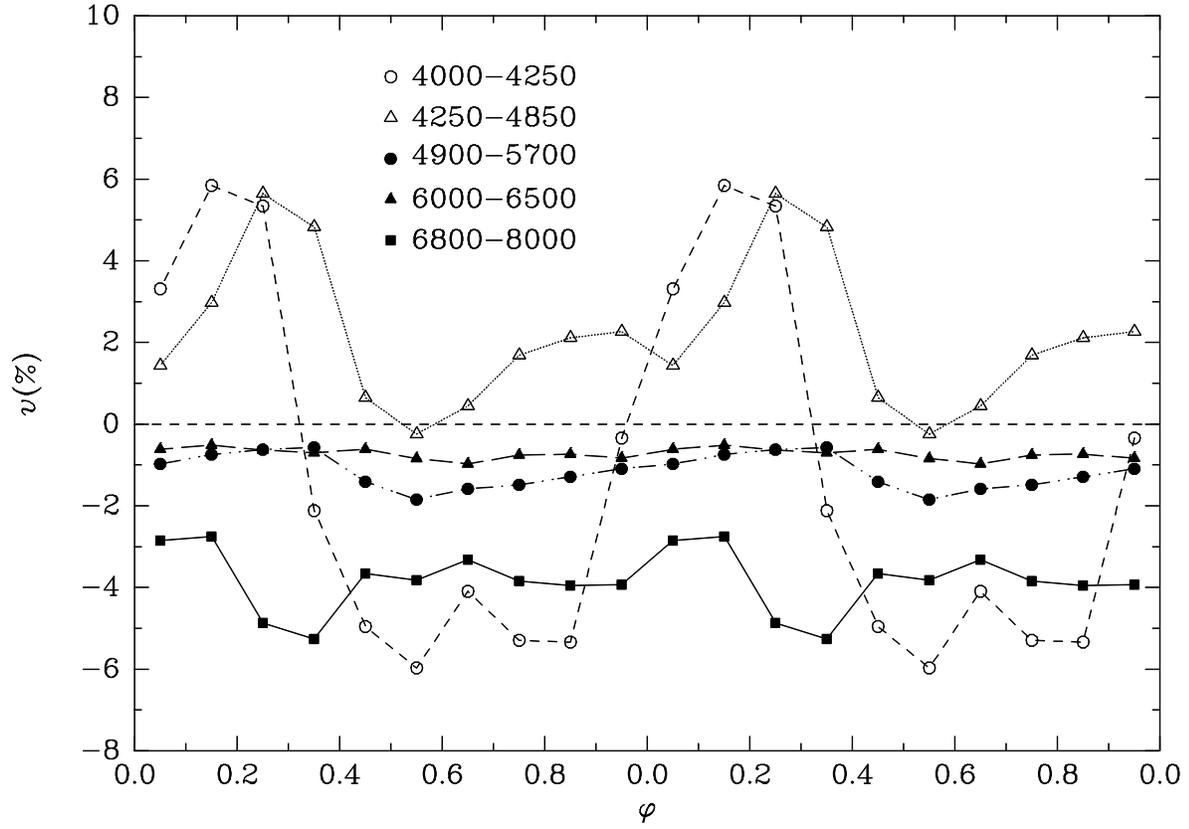}

\caption{The phase dependence of circular polarization in broad spectral bands
chosen to isolate major features.  Note the small and relatively constant
level of polarization in the red.}
\end{figure}

\clearpage

\begin{figure}%Figure 4
\vskip 6.3truein
\includegraphics{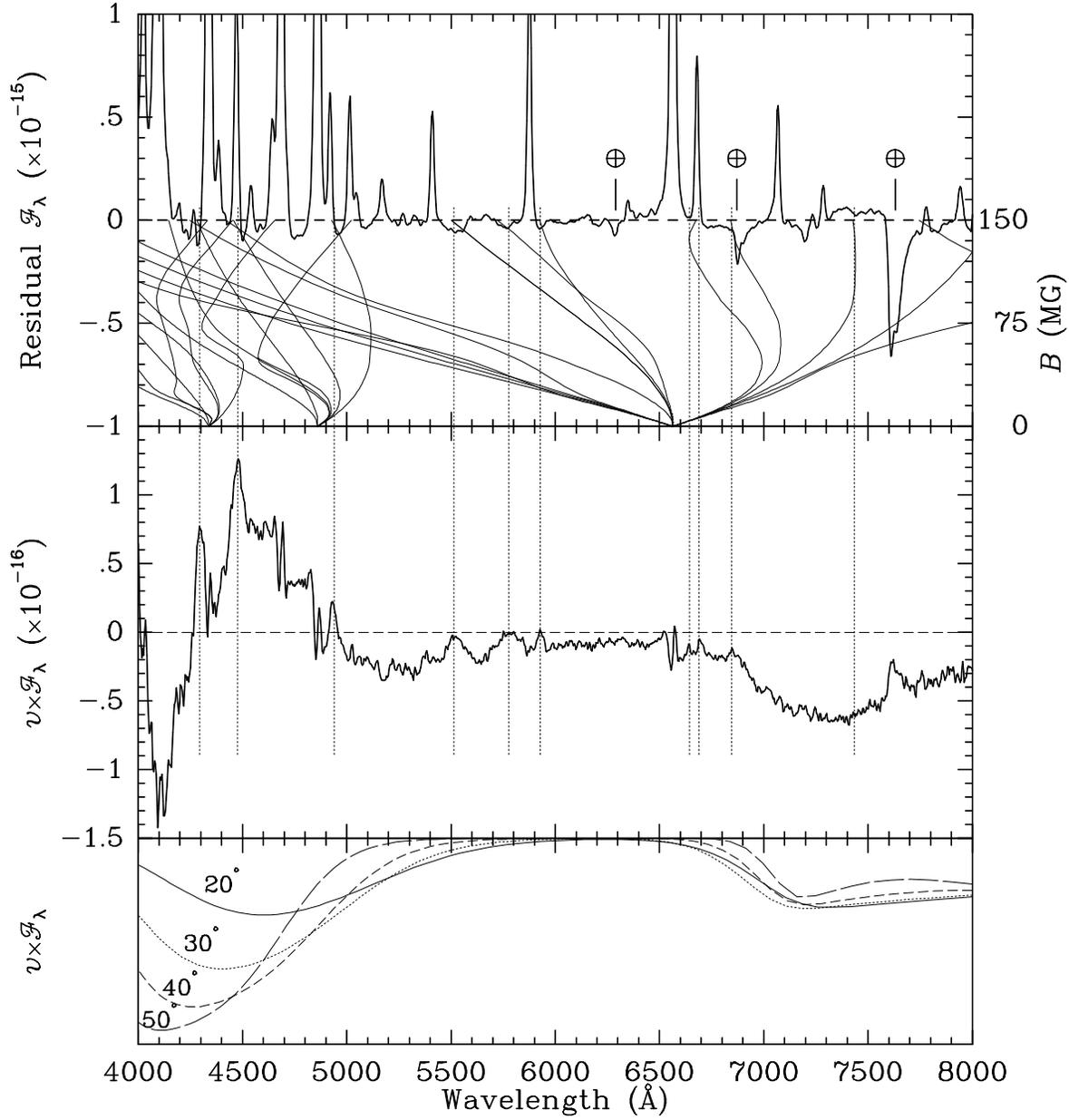}

\caption{{\it Top:\/} Total flux spectrum of \vher\ after subtraction of a
low-order polynomial fit to the continuum.  Smooth lines depict the magnetic
field dependence of the principal lines of hydrogen for
$B=0\rightarrow150$~MG. {\it Middle:\/} Circularly polarized flux coadded over
all orbital phases and epochs.  With orbital motion removed, sharp
polarization reversals are evident across each major emission line, a result
of Zeeman splitting within the funnel.  Several positive-going polarization
features -- indicated by dotted lines -- are accompanied by weak absorption
dips in the total flux, and match the locations of strong hydrogen transitions
for $B=150$~MG. {\it Bottom:\/} Model results for constant-$\Lambda$,
$T_e=5$~keV shocks for the indicated values of $\theta$ from the field
direction.  In order to fix the cyclotron fundamental at its observed
wavelength around 7150~\AA, the field strength varies over
$B=115,120,125,130$~MG for the
$\theta=20\arcdeg,30\arcdeg,40\arcdeg,50\arcdeg$ curves, respectively.}
\end{figure}

\clearpage

\begin{figure}%Figure 5
\vskip 6.8truein
\includegraphics{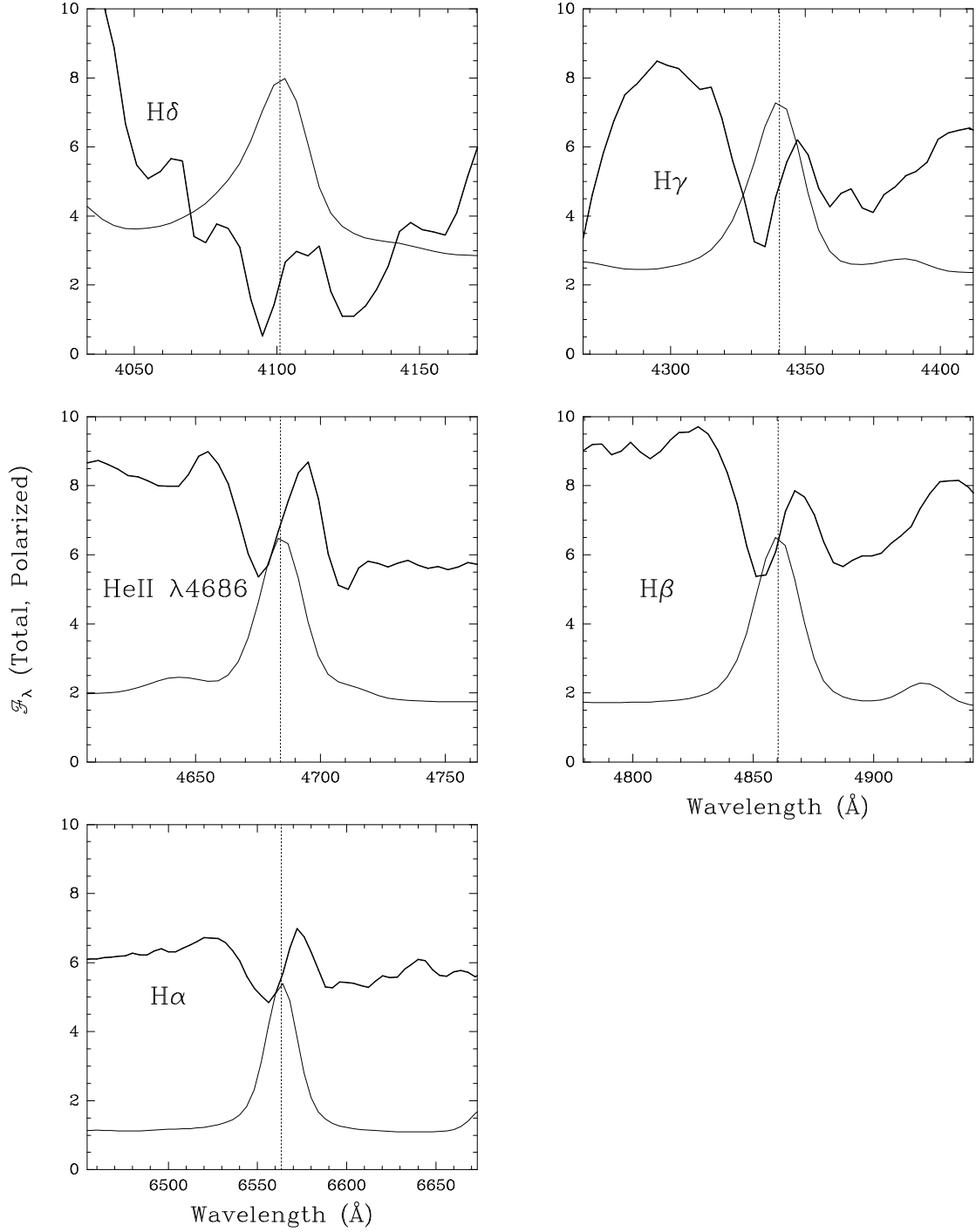}

\caption{Circularly polarized flux {\it (bold)} and total flux {\it (thin)} of
principal emission lines in \vher\ showing Zeeman splitting of the funnel gas
in a longitudinal magnetic field of strength $B_{\rm eff}\approx30$~kG.}
\end{figure}

\clearpage

\begin{figure}%Figure 6
\vskip 5.5truein
\includegraphics{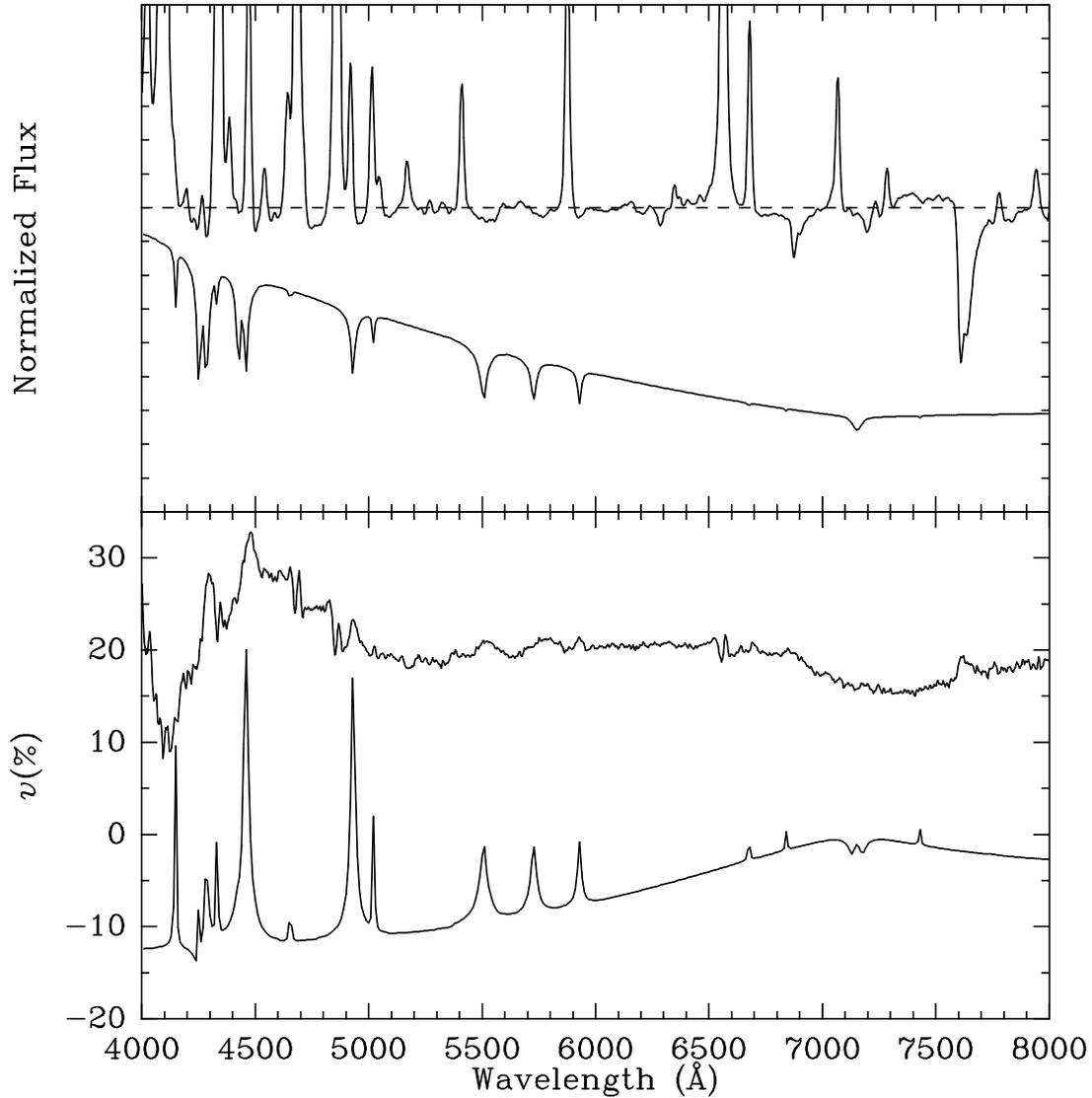}

\caption{Calculated total flux {\it (top panel)\/} and circular polarization
{\it (bottom panel)\/} spectra for a photosphere threaded by a uniform magnetic
field of $B=150$~MG and viewed from an angle of 50\arcdeg\ to the field
direction.  Note the weakness of features in the red and the appearance of the
$\pi$ absorption features between 5500 and 6000~\AA\ in circular polarization,
a result of magneto-optical (Faraday) effects.  The broad hump centered around
$\lambda$7150 is the cyclotron fundamental.  The model curves are compared
with the observed residual flux and circularly polarized flux from Figure~4,
displaced for clarity.}
\end{figure}

\end{document}